\documentclass{aa} 

\IfFileExists{srcltx.sty}{\usepackage[active]{srcltx}}

\usepackage{graphicx}
\usepackage{txfonts}
\usepackage{xspace,natbib,paralist} %
\usepackage{hyperref}
\newcommand{\fov}{{\mathrm{fov}}} 
\newcommand{\vir}{{\mathrm{vir}}} 
\newcommand{\dm}{{\textsc{dm}}} 
\newcommand{\nfw}{{\textsc{nfw}}} 
\newcommand{\iso}{{\mathrm{iso}}} 
\newcommand{\cm}{\:\mathrm{cm}} 
\newcommand{\kev}{\ensuremath{\:\mathrm{keV}}\xspace} 
\newcommand{\kpc}{\:\mathrm{kpc}} 
\newcommand{\parfrac}[2]{\left(\frac{#1}{#2}\right)}

\begin{document}

   \title{Constraints on the parameters of radiatively decaying dark matter 
    from the dark matter halos of the Milky Way and Ursa Minor}

   \author{A. Boyarsky
          \inst{1,2,3}
          \and
           J. Nevalainen
           \inst{4}
           \and
           O.Ruchayskiy
           \inst{5,2}
          }

   \offprints{A. Boyarsky}

\institute{CERN, PH-TH, CH-1211 Geneve 23, Switzerland\\
 \email{alexey.boyarsky@epfl.ch}
\and
\'Ecole Polytechnique F\'ed\'erale de Lausanne, Institute of
  Theoretical Physics, FSB/ITP/LPPC, BSP 720, CH-1015, Lausanne, Switzerland
\and
\emph{On leave of absence from} Bogolyubov Institute of  Theoretical Physics, Kyiv, Ukraine
\and
Helsinki University Observatory, Finland\\
\email{jnevalai@astro.helsinki.fi}
\and
Institut des Hautes \'Etudes Scientifiques, Bures-sur-Yvette,  F-91440, France \\
\email{ruchay@ihes.fr}
}

   \date{Received ; accepted }

  \abstract
   {}
    {We improve the  earlier restrictions on parameters of the dark
    matter (DM) in the form of a sterile neutrino.}
  {The results were obtained from non-observing the DM decay line in the X-ray
    spectrum of the Milky Way (using the recent \emph{XMM-Newton} PN blank sky
    data).  We also present a similar constraint coming from the recent
    \emph{XMM-Newton} observation of Ursa Minor -- dark, X-ray quiet dwarf
    spheroidal galaxy.}
  {The new Milky way data improve on (by as much as the order of magnitude at
    masses $\sim 3.5$ keV) existing constraints.  Although the observation of
    Ursa Minor has relatively poor statistics, the constraints are comparable
    to those recently obtained using observations of the Large Magellanic
    Cloud or M31. This confirms a recent proposal that dwarf satellites of the
    MW are very interesting candidates for the DM search and dedicated studies
    should be made to this purpose.}
   {}


\titlerunning{Dark matter constraints from the Galaxy and Ursa Minor}

   \maketitle


\section{Introduction}

This past year has seen a lot of activity, devoted to searching for the decay
signals of the DM particle in X-ray spectra of various astrophysical
objects~\citep[e.g]{Boyarsky:05,Boyarsky:06b,Boyarsky:06c,Riemer:06,
  Watson:06,Riemer:06b}.  It was noticed long ago by~\cite{Dodelson:93} that a
right-handed neutrino with masses in the keV range presents a viable
\emph{warm dark matter} (WDM) candidate.  Such a particle possesses a specific
radiative decay channel, so one can search for its decay line in the X-ray
spectra of astrophysical objects~\citep{Dolgov:00,Abazajian:01b}.

Recently, the interest in the sterile neutrino as a DM candidate has been
greatly revitalized. First, the discovery of neutrino oscillations (see
e.g.~\citet{Strumia:06} for a review) strongly suggest the existence of
right-handed neutrinos. Probably the easiest way to explain the data on
oscillations is by adding several right-handed, or \emph{sterile}, neutrinos,
to the Standard Model.  It has been demonstrated recently in~\cite{Asaka:05a}
and \cite{Asaka:05b} that a simple extension of the Standard Model by three
singlet fermions with masses smaller than the electroweak scale \citep[dubbed
the $\nu$MSM in][]{Asaka:05a} allows accommodation of the data on neutrino
masses and mixings, allows baryon asymmetry of the Universe to be explained,
and provides a candidate for the dark matter particle in the form of the
lightest of the sterile neutrinos~\footnote{The $\nu$MSM does not explain the
  unconfirmed results of the LSND experiment~\citep{LSND}. There are other
  models that try to account for it by introducing a sterile neutrino with the
  mass around 1 eV. There are also models that explain not all, but only some
  of these phenomena (e.g. LSND and DM, but not the baryon asymmetry as e.g.
  in~\citealt{deGouvea:05}) We do not give any review here. We would like to
  stress that, although our work is motivated by the recent results on the
  $\nu$MSM, our method and results do not rely on any particular model.}.

Secondly, \emph{warm} DM with the mass of particle in keV range can ease the
problem of the dark halo structures in comparison with the cold dark matter
scenario~\cite{Bode:00,Goerdt:06}.  By determining the matter power spectrum
from the Lyman-$\alpha$ forest data from SDSS~\cite{Seljak:06} and
\cite{Viel:06} argue that the mass of the DM particles should be in the range
$\gtrsim 10\kev$ ($\gtrsim 14\kev$ in the case of~\citealt{Seljak:06}). As
this method gives direct bounds for the free-streaming length of the
neutrinos, the bounds on the mass of the DM particle depend on the momentum
distribution function of the sterile neutrinos and, therefore, on their
production mechanism. The results quoted above are claimed for the simplest
Dodelson-Widrow model~(\citeyear{Dodelson:93}).

At the same time, studies of the Fornax dwarf spheroidal
galaxy~\citep{Goerdt:06,Strigari:06} disagree with the predictions of CDM
models and suggest lower mass than in~\cite{Seljak:06} and \cite{Viel:06} for
the DM particle $M_\dm \sim 2\kev$. This result agrees with the earlier
studies of~\citet{Hansen:01} and~\citet{Viel:05}, which used a different
dataset.  For other interesting applications of the sterile neutrinos with the
mass $\sim$ few keV see e.g.~\citet{Kusenko:06a}, \citet{Biermann:06},
\citet{Stasielak:06}, \cite{Kusenko:06b}, and \citet{Hidaka:06}.

It has been argued in~\cite{Boyarsky:06c} and \cite{Riemer:06} that the
preferred targets for observations are objects from the local halo, including
our own Milky Way and its satellites.  In particular,~\cite{Boyarsky:06c}
showed that the best observational targets are the dwarf spheroidals (Ursa
Minor, Draco, etc). Indeed, these objects are X-ray quiet, while at the same
time one expects the DM decay signal from them, comparable to what comes from
galaxy clusters. Because at the time of writing of~\cite{Boyarsky:06c} no
public data were available for these dwarf spheroidals, the observations of
the core of Large Magellanic Cloud (LMC) were used to produce the strongest
restrictions on parameters of the sterile neutrino.  It was stressed
in~\cite{Boyarsky:06c} that other dwarf satellite galaxies should be studied
as well, in order to minimize uncertainties related to the DM modeling in each
single object.  In this paper we continue studies of the dwarf satellites of
the MW by analyzing the data from \emph{XMM-Newton} observation of
Ursa Minor and confirm the restrictions of~\cite{Boyarsky:06c}.

It was also shown in~\cite{Boyarsky:06c} that the improvement of the results
from MW DM halo can be achieved by using longer exposure data (notably, longer
exposure of the closed filter observations). In this paper, we improve our
restrictions, coming from the MW DM halo by using the blank sky dataset with
better statistics from~\cite{Nevalainen:05}.

\section{DM with radiative decay channel}
\label{sec:properties}

Although throughout this paper we are talking mostly about the sterile
neutrino DM, the results can be applied to \emph{any} DM particle that
possesses the monoenergetic radiative decay channel, emits photon of energy
$E_\gamma$ and has a decay width $\Gamma$. In the case of the sterile neutrino
(with mass below that of an electron), the radiative decay channel is into a
photon and active neutrino \cite{Pal:81}. As the mass of an active neutrino is
much lower than keV,  $E_\gamma = \frac{M_s}2$ in this case. The width $\Gamma$
of radiative decay can be expressed \citep{Pal:81,Barger:95} in terms of mass
$M_s$ and \emph{mixing angle} $\theta$ via
\begin{equation}
  \label{eq:1}
  \Gamma = \frac{9\, \alpha\, G_F^2} {1024\pi^4}\sin^22\theta\, M_s^5  \simeq 1.38\times10^{-22}\sin^2(2\theta)
  \left[\frac{M_s}{\mathrm{keV}}\right]^5\;\mathrm{sec}^{-1}\:.
\end{equation}
(The notation $\sin^2(2\theta)$ is used traditionally, although in all
realistic cases $\theta\ll 1$).  The flux of the DM decay from a given
direction is given by
\begin{equation}
\label{eq:3}
  F_{\dm} =  \Gamma\frac{E_\gamma}{M_s}\int\limits_{\fov\;
    \mathrm{cone}}\frac{\rho_\dm(\vec{r})}{4\pi|\vec D_L + \vec r|^2}d^3\vec
  r\;.
\end{equation}
Here $\vec D_L$ is the \emph{luminous} distance between the observer and the
center of the observed object, $\rho_\dm(r)$ is the DM density, and the
integration is over the DM distribution inside the (truncated) cone -- solid
angle, spanned by the field of view (FoV) of the X-ray satellite.  If the
observed object is far,\footnote{Namely, if luminosity distance $D_L$ is much
  greater than the characteristic scale of the DM distribution $\rho_\dm(r)$.}
then Eq.~(\ref{eq:3}) can be simplified:
\begin{equation}
\label{eq:4}
F_\dm = \frac{M_\dm^\fov \Gamma}{4\pi D_L^2}\frac{E_\gamma}{M_s}\;,
\end{equation}
where $M_\dm^\fov$ is the mass of DM within a telescope's field of view (FoV).
Equation~(\ref{eq:4}) can be rewritten again as
\begin{equation}
  \label{eq:5}
  F_{\dm} = 6.38 \times 10^6\parfrac{M_{\dm}^\fov}{10^{10}M_\odot}\parfrac{\!\kpc}{D_L}^2 
  \times \sin^2(2\theta) \left[\frac{M_s}{\mathrm{keV}}\right]^5   \frac{\mathrm{keV}}{\mathrm{cm^2 \cdot sec}}.
\end{equation}
In the absence of a clearly detectable line, one can put an upper limit on the
flux of DM from the astrophysical data, which  will lead via Eq.~(\ref{eq:5})
to the restrictions of parameters of the sterile neutrino $M_s$ and $\theta$.

\section{Restrictions from the blank sky observation}
\label{sec:blank-sky}

\subsection{Modeling the DM halo of the MW}
\label{sec:mw}

As shown in the previous section, one needs to know the distribution of the DM
to obtain the restrictions on parameters of the sterile neutrino. In the case
of nearby objects (including our own Galaxy and dwarf satellites from the
local halo), the DM distribution can be deduced e.g. by using the rotation
curves of the stars in the galaxy. Here we follow the analysis
of~\cite{Boyarsky:06c}.  Various DM profiles, used to fit observed velocity
distributions, differ the most in the center of a distribution. In the case of
the MW we choose, as in~\cite{Boyarsky:06c}, to use the observations away from
the center, to minimize this uncertainty.  In particular, in
Refs.~\cite{Klypin:02,Battaglia:05} it was shown that the DM halo of the MW
can be described by the Navarro-Frenk-White (NFW) profile~\citep{Navarro:96}
\begin{equation}
  \label{eq:6}
  \rho_\nfw(r) = \frac{\rho_s r_s^3}{r(r+r_s)^2}\;,
\end{equation}
with parameters, given in Table~\ref{tab:nfw}.\footnote{According
  to~\cite{Klypin:02}, the choice of e.g. the Moore profile~\cite{Ghigna:99}
  or a generalization thereof, as compared to the NFW profile, would change
  the results by $\lesssim 1\%$ for $r< 3\kpc$. As we are using observations
  away from the center, this difference is completely negligible, so we
  choose to use the NFW profile.} %
The relation between virial parameters and $\rho_s$, $r_s$ are given in the
Appendix~\ref{sec:param-nfw}.  Quoted halo parameters provide DM decay flux
(from the directions with $\phi>90^\circ$) consistent within $\sim 5\%$ with
the one, given by Eqs.~(\ref{eq:10})--(\ref{eq:19}). Only ``maximal disk''
models in \cite{Klypin:02} would provide $30-50\%$ weaker restrictions;
however, these models are highly implausible, see \cite{Klypin:02}. Similarly,
taking the lower limit for the virial mass of \cite{Battaglia:05}, one would
obtain 25\% weaker restrictions than the ones, presented in this
paper.\footnote{When quoting results of~\cite{Klypin:02}, we do not take the
  effects of baryon compression on DM into account. While these effects make
  DM distribution in the core of the MW denser, they are hard to compute
  precisely.  Thus the values we adopt give us a conservative lower bound on
  the estimated DM signal.}

%
\begin{table*}
\caption{Best-fit parameters of NFW model of the MW DM halo.}
\label{tab:nfw}
\centering
\begin{tabular}{c c c c c c}     
\hline\hline
References & $M_\vir$ [$M_\odot$] & $r_\vir$ [kpc] & Concentration & $r_s$ [kpc] & $\rho_s$ [$M_\odot/\mathrm{kpc}^3$]\\
\hline
\cite{Klypin:02}, favored models ($A_1$ or $B_1$) & $1.0\times 10^{12}$ & 258 & 12 & 21.5 & $4.9\times 10^6$\\
\cite{Battaglia:05}                               & $0.8^{+1.2}_{-0.2}\times 10^{12}$  & 255 &    18 & 14.2 & $11.2\times 10^6$\\
\hline
\end{tabular}
\end{table*}
%

To compare the results from different (e.g. cuspy and cored) profiles, we can
also describe the DM distribution in the MW via an isothermal profile:
\begin{equation}
  \label{eq:10}
  \rho_\iso(r) =\frac{v_h^2}{4\pi G_N}\frac1{r^2 + r_c^2}\;.
\end{equation}
The DM flux from a given direction $\phi$ into the solid angle $\Omega_\fov
\ll 1$, measured by an observer on Earth (distance $r_\odot\simeq 8\kpc$ from
the galactic center), is given by
\begin{equation}
  \label{eq:19}
    F_\dm^\iso(\phi)\, {=}\, \frac {L_0}{R} \times \left \{ 
    \begin{array}{ll}
     \!\!\frac {\pi}{2} + 
        \arctan\left(\frac{r_\odot\cos\phi}
        {R}\right) , &   \cos\phi \ge 0 \\
         \!\arctan\left(
  \frac{R}{r_\odot|\cos\phi|}\right), & \cos\phi < 0
        \end{array} \right. .
\end{equation}
Here $L_0\equiv\frac{\Gamma \Omega_\fov v_h^2}{32\pi^2 G_N}$ and
$R=\sqrt{r_c^2 + r_\odot^2\sin^2\phi}$. Angle $\phi$ is related to the
galactic coordinates $(b,l)$ via
\begin{equation}
\label{eq:11}
\cos\phi =\cos b\,\cos l\;.
\end{equation}
Thus, the galactic center corresponds to $\phi=0^\circ$, and the anti-center
$\phi=180^\circ$ and the direction perpendicular to the galactic plane to
$\phi=90^\circ$.

In~\cite{Boyarsky:06c}, the following parameters of isothermal profile were
chosen: $v_h = 170$~km/sec and $r_c = 4\kpc$.  One can easily check (using
Table~\ref{tab:nfw} and Eqs.~(\ref{eq:13})--(\ref{eq:14}) in
Appendix~\ref{sec:param-nfw}) that, in the directions $\phi\gtrsim 90^\circ$,
the difference in predicted DM fluxes between the NFW model with parameters,
given in Table~\ref{tab:nfw} and isothermal model with parameters just quoted
are completely negligible (less than 5\%).\footnote{Ref.~\cite{Abazajian:06}
  claim that the MW results of Refs.~\cite{Boyarsky:06c,Riemer:06} are
  uncertain by about a factor of 3.  This conclusion was based on the range of
  virial masses of the MW DM halo $M_\vir = (0.7-2.0)\times 10^{12}\:M_\odot$
  in Ref.~\cite{Klypin:02}.  However, as just demonstrated, the authors
  of~\cite{Boyarsky:06c} have chosen parameters of DM halo conservatively. The
  flux they used, corresponded to the favored models $A_1$ or $B_1$
  in~\cite{Klypin:02}, with $M_\vir\sim 1.0\times 10^{12}M_\odot$. These
  models provide the \emph{lowest} bound on the derived flux of DM decay (if
  one does not take into account the highly implausible ``maximum disk''
  ($A_2$ or $B_2$) models of~\cite{Klypin:02}).  Even in the latter case, the
  DM flux will be $30-50\%$ lower than the one used in
  work~\cite{Boyarsky:06c}.  Therefore, parameters of the MW DM halo
  from~\cite{Boyarsky:06c} provide the conservative estimate so we use them in
  our work as well.}

\subsection{XMM-Newton PN blank sky data}
\label{sec:blank-sky-data}

To examine the Milky Way halo, we used the double-filtered, single+double
event XMM-Newton PN blank sky data from \cite{Nevalainen:05}, which is a
collection of 18 blank sky observations (see Table~2 in \cite{Nevalainen:05}
for their observation IDs, positions, and exposures).\footnote{We processed
  the blank sky data with newer SAS distribution,
  \texttt{xmmsas\_20050815\_1803-6.5.0}, and obtained
  slightly different exposure times than those in the public data.} %
The exposure time of the co-added observations is 547 ks. We used a combination of closed-filter observations
from \cite{Nevalainen:05} (total exposure time 145 ks) to model  the 
background of \emph{XMM-Newton} PN instrument separately.
The data has been filtered using SAS expression ``flag==0'', which rejects the data
from bad pixels and CCD gap regions. After removing the brightest point
sources, the 
total accumulation area is 603 arcmin$^2$.

Based on the $>$10 keV band count rates of the blank sky and the closed filter data, 
we normalized the closed-filter spectrum by a factor of 1.07 before subtracting it from the 
blank sky spectrum. The remaining sky-background spectrum consists mainly of the Galactic emission
and the cosmic X-ray background (CXB) due to unresolved extragalactic point sources. 
We modeled the Galactic emission by a non-absorbed MEKAL model with Solar abundances.
For the CXB emission, we used a power-law model modified at the lowest energies by Galactic absorption with the value of $N_H$
fixed to its exposure-weighted average over all blank sky observations
($N_H=1.3\times 10^{20}\cm^{-2}$). 

\begin{figure}[t]
  \centering \includegraphics[angle=0,width=\columnwidth]{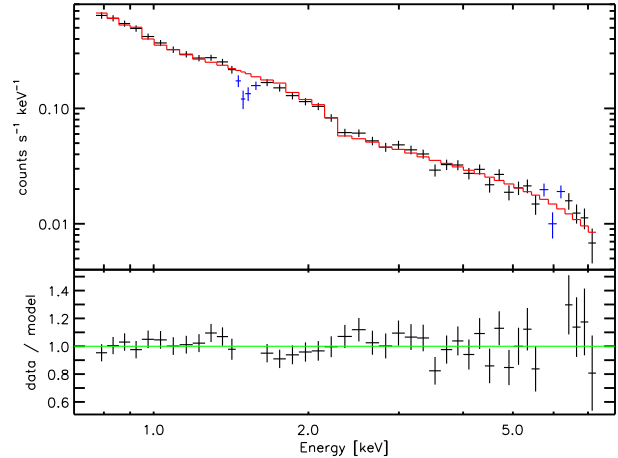}
  \caption{{\it Upper panel:} The  blank sky data of \cite{Nevalainen:05}
    after subtraction of the closed filter data. The black crosses show the
    data points, used for modeling the Galactic emission and CXB. The blue
    crosses show the data points excluded from the fit. The best-fit model is
    shown with the red line.  The plotted spectrum is binned by a minimum of
    20000 counts per channel, different from that used in the actual fit (see
    text).  The error bars include the 5\% systematic uncertainty used in the
    analysis.  {\it Lower panel:} The ratio of the data-to-model values in
    each channel used in the fit. }
  \label{fig:mw-xspec}
\end{figure}

The variable Galactic emission and geocoronal Solar wind-charge exchange
emission (see e.g.~\citealt{Wargelin:04}) complicate the modeling at the
lowest energies.  The remaining calibration uncertainties further complicate
the analysis in the lowest energies (see e.g. \citealt{Nevalainen:06}).  Thus,
we omitted the channels below 0.8 keV.  At energies above 7 keV, the particle
background dominates and the total flux is very sensitive to the background
normalization. We thus excluded channels above 7 keV.

The data are not well-described in the 1.45--1.55 and 5.8--6.3 keV bands with
the above model. These deviations probably originate from the variability of
the instrumental Al and Fe line emission. In order to minimize the effect of
the instrumental problems, we excluded these bands when finding the best-fit
sky background model (see below).  Also,  to account for possibly
remaining calibration inaccuracies, we added a systematic uncertainty of 5\%
of the model value in each bin in quadrature to the statistical uncertainties.

We binned the spectrum using a bin size of 1/3 of the energy resolution and
fitted the data using models and channels as described above.  The best-fit
(reduced $\chi^{2}$ = 1.03 for 153 degrees of freedom) model agrees with the
data within the uncertainties (see Fig. \ref{fig:mw-xspec}), yielding a photon
index of 1.50$\pm$0.02 at 1 $\sigma$ confidence level (see
Fig.\ref{fig:mw-xspec}), consistent with  similar analyses
based on \emph{Chandra} \citep{Hickox:06} and XMM-Newton MOS instrument
\citep{DeLuca:03}. The best-fit temperature of the MEKAL component used to
model the Galactic emission is 0.19$\pm$0.01 keV, consistent with
e.g.~\cite{Hickox:06}.

We then evaluated the level of possible DM flux above the background model
allowed by the statistical and systematic uncertainties in each channel.  For
this, we modified the above best-fit model by adding a narrow (width = 1 eV)
Gaussian line to it.  We then re-fitted the data, fixing the Gaussian centroid
for each fit to the central energy of a different channel.  In these fits we
fixed the above continuum model parameters to the best-fit values and thus the
Gaussian normalization parameter is the only free parameter. We used the fits
to find the upper 3$\sigma$ uncertainty of the Gaussian normalization, i.e.
the allowed DM flux.  Note that here we included the channels 1.45--1.55 and
5.8--6.3 keV (excluded above when defining the sky background model).  The
background is oversubtracted in the channels at 1.45-1.55 keV and $\sim$6.0
keV (see Fig.~\ref{fig:mw-xspec}), which would formally require negative
normalization for the Gaussian. However, we forced the normalization to be
positive and thus obtained conservative upper limits in these energies.

\begin{figure}[t]
  \centering
  \includegraphics[width=\columnwidth]{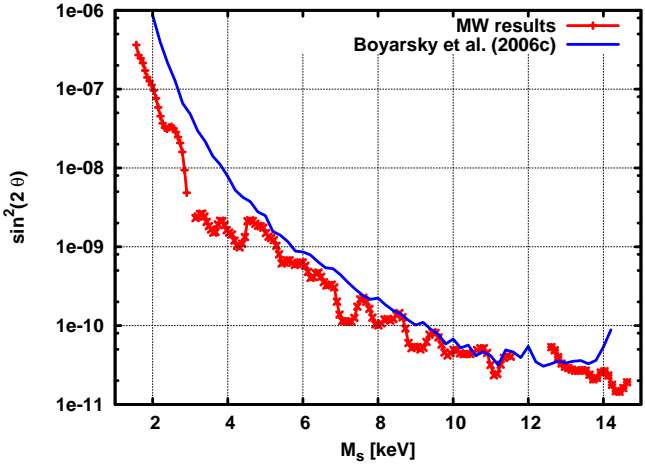}%
  \caption{Exclusion plot based on the blank sky observations. 
    Bins $2.9-3.1\kev$ and $11.6-12.6\kev$ are excluded.}
  \label{fig:mw}
\end{figure}

Finally, we converted the  upper bound obtained for the flux per energy bin to the
restrictions on $M_s$ and $\sin^22\theta$, using
Eqs.~(\ref{eq:1}),~(\ref{eq:19}) (we use exposure weighted average of DM
fluxes~(Eq. \ref{eq:19}) from all the observations, constituting the blank sky
dataset). This corresponds to the average ``column density'' $1.22 \times
10^{28}$ keV/cm$^2$.  The results are shown in Fig.~\ref{fig:mw}
%

%

\begin{figure}[t]
  \centering
  \includegraphics[width=\columnwidth]{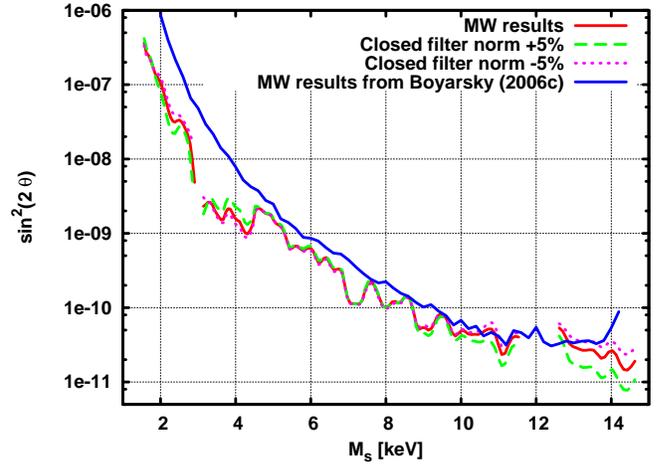}%
  \caption{Dependence of the results on closed filter normalization. Red
    (solid) and green (long-dashed) lines are the same as in
    Fig.~\ref{fig:mw}.}
  \label{fig:closed}
\end{figure}

At energies above $E=5$ keV, the instrumental background of PN dominates over
the sky background \citep[c.f.][]{Nevalainen:05}. Therefore, the accuracy of
the co-added closed filter spectrum in predicting the particle background in
the blank sky observations becomes essential.  We estimate this accuracy using
the variability of the individual closed-filter spectra in the 0.8-7.0~keV
band \citep{Nevalainen:05} and propagate it by varying the normalization of
closed filter data by $\pm 5\%$ and repeating the above analysis.  This leads
to a factor of 3 change in the results at $M_s \sim 14$ keV (see
FIG.~\ref{fig:closed}). Therefore, for $E\gtrsim 5$ keV we choose the more
conservative normalization (see FIG.~\ref{fig:results} below).

%

\section{Restrictions from observations of Ursa Minor}
\label{sec:ursa}

It was argued in \cite{Boyarsky:06c} that dwarf satellite galaxies should
provide the best restrictions, based on their high concentration of DM and low
X-ray signal. At the moment of writing of \cite{Boyarsky:06c}, no public data
on preferred dwarf satellites were available, therefore the observation of
core of LMC were used as a demonstration. Recently, the Ursa Minor dwarf (UMi)
was observed with \emph{XMM-Newton} (obs IDs.: 0301690201, 0301690301,
0301690401, 0301690501, observed in August-September 2005).\footnote{We are
  very grateful to Prof. T.~Maccarone for sharing this data with us before it
  became publicly available through the XMM data archive.} %
Unfortunately, most of these observations are strongly contaminated by
background
flares and the observations have very small exposure times.  Below we present the
analysis of only one observation (obsID: 0301690401), which ``suffered'' the
least from background contamination.

\subsection{DM modeling for UMi}
\label{sec:dm-ursa}

The DM distribution in UMi has a cored profile~\citep[see
e.g.][]{Kleyna:03,Wilkinson:06,Gilmore:06,Gilmore:07}.\footnote{As discussed
  in Section~\ref{sec:mw}, the estimates for DM flux do not vary significantly
  if one uses NFW instead of the isothermal DM density profile. In the case of
  UMi, 
  the cored (isothermal) profile will clearly produce a more conservative
  estimate than will the cuspy NFW profile. Indeed, taking NFW parameters for UMi from
  the recent paper~\cite{Wu:07} gives a $\sim 20\%$ higher
  estimate for the DM mass within the FoV.} %
We adopt the following parameters of isothermal profile~(\ref{eq:10}) for UMi:
$v_h = 23$~km/sec, $r_c = 0.1\kpc$~(see
e.g.~\citealt{Wilkinson:06}).\footnote{For the detailed studies of mass
  distribution in dwarf spheroidals, see~\cite{Gilmore:07}. We are grateful to
  Prof. G.~Gilmore for sharing the numbers with us before their paper became
  available.  The statistical uncertainty in determining  these numbers is
  below 10\%.  The systematic uncertainties are much harder to estimate. One
  of the major sources of the systematic errors comes from violation of the
  main assumptions of the method: deviation from equilibrium and from the
  spherical distribution of matter in a galaxy.  In other known examples it
  provides a factor of 2 uncertainty, which should be a conservative estimate
  in the case of UMi, as it is rather spherical. Another typical uncertainty
  -- determination of the mass of the stars -- is not important for UMi, as it
  has a very high mass-to-light ratio.}
We adopt the distance to UMi
$D_L = 66\kpc$ \citep{Mateo:98}. The DM mass within the circular FoV with the
radius $r_\fov$, centered at the center of the galaxy is given by
\begin{equation}
  \label{eq:15}
  M_\dm^\fov = \frac{\pi v_h^2}{2 G_N}\left(\sqrt{r_\fov^2+r_c^2}-r_c\right)\;.
\end{equation}
In our case, the radius of FoV is 13.9', which corresponds to $r_\fov =
0.27\kpc$ (i.e. about $3r_c$). Therefore
\begin{equation}
  \label{eq:16}
 M_{\dm}^\fov =3.3\times 10^7 M_\odot \ \ {\rm for} \ \  r_\fov = 0.27\kpc .
\end{equation}
Using Eqs.~(\ref{eq:16}) and (\ref{eq:4}), one can compute the expected DM
flux from UMi:
\begin{equation}
  \label{eq:17}
  F_{\dm} = 4.79 \ \frac{\mathrm{keV}}{\mathrm{cm^2\cdot      sec}}  \left[\frac{M_s}{\mathrm{keV}}\right]^5  \sin^2(2\theta) .
\end{equation}

\subsection{PN data analysis}
\label{sec:background}

%
\begin{figure}[t]
  \centering
  \includegraphics[width=\columnwidth]{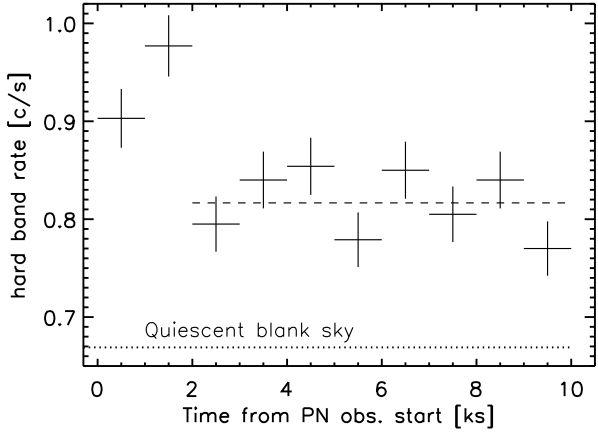}
  \caption{Hard band light curve for the UMi observation 0301690401. The crosses show the PN $>$10 keV band rate of the full FOV 
    in 1ks bins. The dashed line shows the average, when excluding first 2
    time bins. The dotted line shows the corresponding quiescent value in the
    co-added blank sky data \citep{Nevalainen:05}. }
  \label{fig:light-curve}
\end{figure}
%

We processed the Ursa Minor observation 0301690401 using \texttt{epchain}
version 8.56 and filtered the event file with SAS expressions
``\texttt{PATTERN<=4}'' and ``\texttt{FLAG==0}''.  We applied the blank sky-based XMM-Newton background method of \cite{Nevalainen:05} for Ursa Minor.
The $>$ 10 keV band light curve from the full FOV (FIG.~\ref{fig:light-curve})
shows that the count rate in observation 0301690401 (excluding first 2ks)
exceeds that of the blank sky quiescent average by 25\%.  This level is
higher, but close to what is used in the blank sky accumulation ($\pm$ 20\%
filtering around the quiescent level).  Thus we accepted the data from all
instants after the initial 2ks, and we approximated the background
uncertainties with those in \cite{Nevalainen:05}.

The hydrogen column density in the direction of Ursa Minor is small ($\sim$$
N_H=2\times 10^{20}\cm^{-2}$) and consistent with the variation in the blank
sky sample.  Thus we can also apply the blank sky background method  to
channels below 2 keV.

As noted in the above XMM-Newton blank sky study, the $>$ 10 keV band-based
scaling of the background only works up to a factor of 1.1, beyond which the
background prediction becomes worse. Furthermore, the correlation of
background rates in the $>$ 10 keV band is very poor with the rates below 2
keV band. Thus, in order to achieve the best possible background prediction
accuracy, we scaled the blank sky background spectrum by a factor of 1.1 at
channels above 2 keV, and  we applied no background scalingat lower energies.

We removed this scaled background spectrum from the Ursa Minor spectrum (see
FIG.~\ref{fig:flux-umi}).  As shown in \cite{Nevalainen:05}, the background
accuracy is worse at lower energies.  We used those estimates to propagate the
background uncertainties at 1 $\sigma$ confidence level to our results by
examining how the results change when varying the 0.8--2.0 keV and 2.0--7.0
keV band background by 15\% and 10\%, respectively.

\subsection{Ursa Minor data and restrictions on the sterile neutrino parameters}
\label{sec:ursa-data}

\begin{figure}[t]
  \centering \includegraphics[width=\columnwidth]{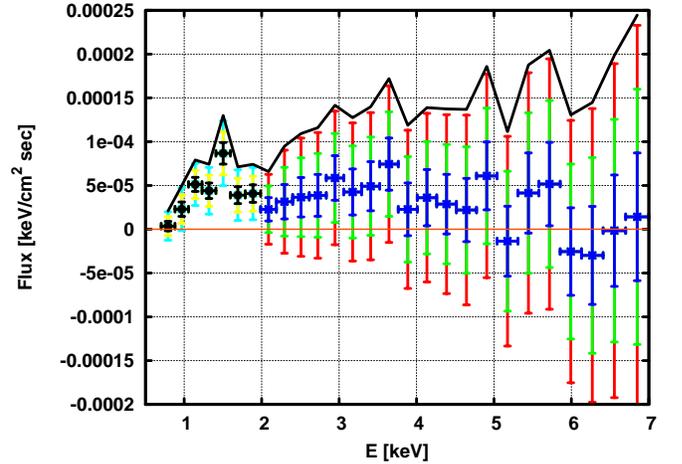}
  \caption{Flux from UMi (obs. ID 0301690401). Energy bins have the width of
    twice the spectral resolution. Shown are the 1, 2, and 3$\sigma$ errors.
    One can see that, above 2 keV, flux in most energy bins is zero within
    $1\sigma$ limits (blue crosses) and for the rest it is zero within
    $2\sigma$ limits (green crosses). Similarly, below 2 keV black, cyan and
    yellow crosses represent 1,2, and 3$\sigma$ error correspondingly.  The
    solid black line represents the 3$\sigma$ upper bound on total flux in a
    given energy bin, which we use to put the limit on DM parameters.}
  \label{fig:flux-umi}
\end{figure}

\begin{figure}[t]
  \centering %
  \includegraphics[width=\columnwidth]{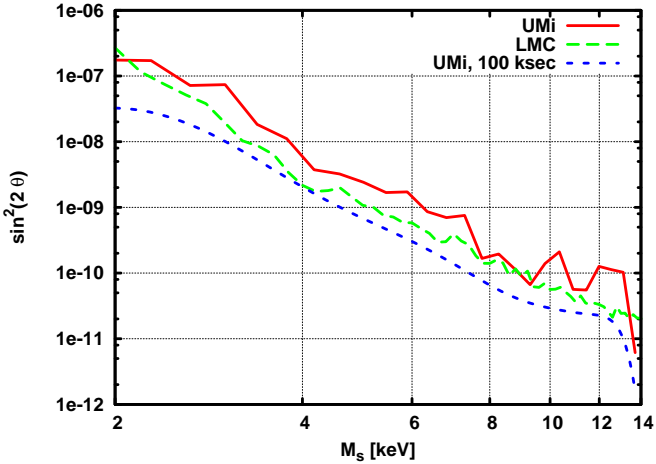}
  \caption{Exclusion from UMi (red solid line), as compared to LMC (green
    long-dashed line). The blue short-dashed smooth curve shows hypothetical
    restrictions from UMi observations with 100 ksec exposure.}
  \label{fig:ursa}
\end{figure}

The X-ray spectrum of UMi is similar to that of LMC: above 2 keV the flux is
zero within statistical limits (see FIG.~\ref{fig:flux-umi}). (Of course, the
data set has rather low statistics: after the cleaning of flares the UMi
observation only contains 7 ks).
%
%
Therefore, for such data, we utilized the ``total flux'' method. Namely, we
restricted the DM flux in the given energy bin to be bounded from above by the
measured total flux in this energy bin plus its $3\sigma$ uncertainty
(FIG.~\ref{fig:flux-umi}). As each energy bin has a width of twice the FWHM at
a given energy, the flux from a DM line would not ``spill'' into nearby bins.
Using Eq.~(\ref{eq:17}), we find the restrictions on the sterile neutrino
parameters, represented in FIG.~\ref{fig:ursa}.
%
%

These restrictions should be compared with those, obtained from another
satellite galaxy -- Large Magellanic Cloud~\citep{Boyarsky:06c}. As one
clearly sees from FIG.~\ref{fig:ursa}, in spite of the low exposure time, it
is fully consistent with the earlier bounds from LMC, thus confirming the
results of~\cite{Boyarsky:06c}. Improvement of the exposure for UMi
observations should, presumably lead to the improvement of results (at least
for energies above $E\gtrsim 2$ keV). For example, for a 100 ksec observation,
we expect the results to improve by roughly a factor
$\sqrt{100\:\mathrm{ks}/7\:\mathrm{ks}}\approx 3.77$.

\section{Results}
\label{sec:results}

\begin{figure}[t]
  \centering \includegraphics[width=\columnwidth]{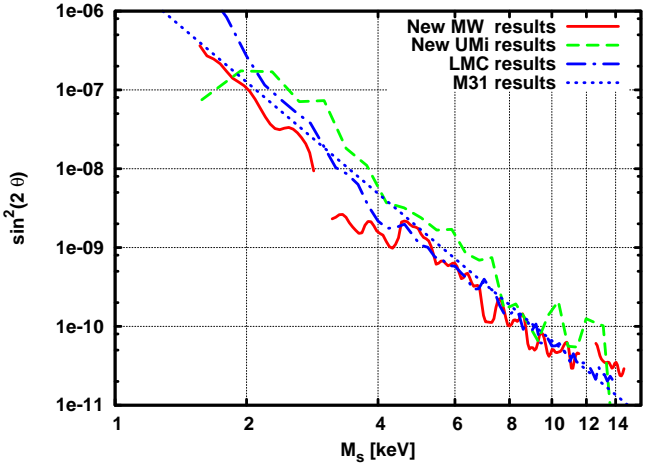}
  \caption{Results and comparison with previous bounds (a region of parameter
    space \emph{above} a curve is excluded)}
  \label{fig:results}
\end{figure}

\subsection{Restrictions from the blank sky data}
\label{sec:blank-sky-res}

By analyzing the blank sky data set with better statistics, we improved on the
previous results of \cite{Boyarsky:06c}, \cite{Riemer:06}, \cite{Watson:06}
(by as much as the factor of 10 for $M_s \approx 3.5$ keV and by the
negligible amount for $M_s\gtrsim 11$ keV). The result is shown in
Fig.~\ref{fig:results} in red solid line. The best previous bounds are also
shown: bound from LMC \citep{Boyarsky:06c} with a blue short-dashed line and
bound from M31 \citep{Watson:06} with a dotted magenta line. We see that in the
region 3.5 keV $\lesssim M_s\lesssim 11$ keV the new blank sky data improves
on previous results. These results can be converted (using
Eq.~(\ref{eq:1})) into restrictions on the decay rate $\Gamma$ of \emph{any}
DM particle, that possesses radiative decay channel and emits a photon
$E_\gamma$ (see FIG.~\ref{fig:gamma}).  Our results provide more than an order
of magnitude improvement over similar restrictions derived in \cite{Riemer:06}
(which used the \emph{Chandra} blank sky background), as one can clearly see
by comparing FIG.~\ref{fig:gamma} with FIG.~2 in \cite{Riemer:06}, where
the exclusion plot is above $\Gamma = 10^{-26}\;\mathrm{sec}^{-1}$ line for all
energies. (In \cite{Riemer:06} the restriction were made, based on the total
flux of \emph{Chandra} satellite, without subtraction of the instrumental
background, which explains a much weaker restrictions).

The empirical fit to the MW data is given by the following expression:
\begin{equation}
  \label{eq:18}
  \sin^2(2\theta) \lesssim 2.15\times
  10^{-7}\parfrac{M_s}{\mathrm{keV}}^{-3.45}\;.
\end{equation}

\subsection{Restrictions from Ursa Minor dwarf}
\label{sec:ursa-res}

Restrictions from XMM observation of UMi are shown in FIG.~\ref{fig:results}
by the green long-dashed line. These results are slightly weaker than LMC or
M31 results, which is due to the very low statistics of the UMi observation.
Improvement of the statistics should lead to improvement of the current bound
(as shown on the FIG.~\ref{fig:ursa}). These results confirm the recent claims
(Boyarsky et al., 2006c) that dwarfs of the local halo are promising
candidates for the DM-decay line search and, as such, should be studied
dedicatedly.

In searching for the DM signatures, it is important to understand that the
uncertainties of the DM modeling for any given object can be large, and
therefore it is important to study many objects of given type, as well as many
different types of objects (where DM distributions are deduced by independent
methods). To this end, although UMi data does not provide any improvement over
existing bounds, it makes those bounds more robust as the existence of DM in
UMi is deduced by independent observations, and the rotation curves of UMi are
measured quite well, since it has less perturbed dynamics, compared to e.g.
LMC.

\section{Discussion}
\label{sec:discussion}

%
\begin{figure}[t]
  \centering \includegraphics[width=\columnwidth]{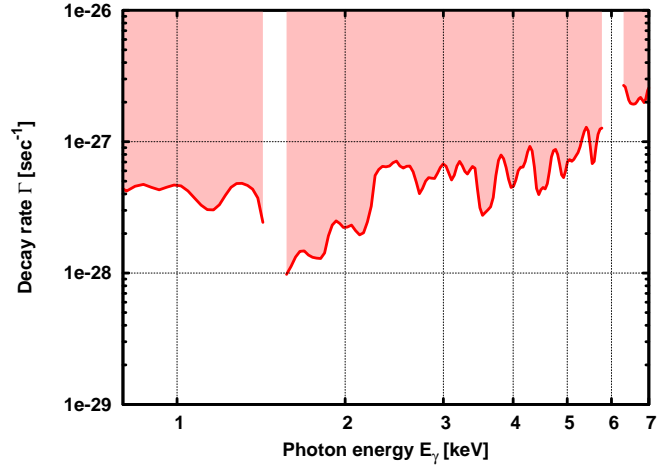}
  \caption{Restrictions on parameters of any DM particle with the radiative
    decay width $\Gamma$, emitting photon of energy $E_\gamma$. Shaded region
    is excluded.}
  \label{fig:gamma}
\end{figure}
%

In this paper we continued to search for the best astrophysical objects, from
the point of view of restricting parameters of DM particles with the radiative
decay channel. Several comments are in order here.

\begin{compactenum}[(1)]
\item Although throughout this paper we have spoken about the sterile neutrinos and
  restricted their parameters (namely, mass, and mixing angle), the constraints
  can be readily converted into any other DM candidate that possesses a
  radiative decay channel. \footnote{For earlier works, discussing
    cosmological and astrophysical effects of decaying DM,
    see e.g.~\cite{DeRujula:80,Berezhiani:87,Doroshkevich:89,Berezhiani:90a,%
      Berezhiani:90b}.  The extensive review of the results can also be found
    in the book~\cite{Khlopov:97}.} Then the restrictions are formulated on
  the decay rate $\Gamma$ as a function of energy of the emitted photon
  $E_\gamma$. The results then can be presented in the form of an exclusion
  plot, presented in FIG.~\ref{fig:gamma}.
  
\item Clearly, if one could relate parameters of the sterile neutrino with
  their relic abundance $\Omega_s$, this would allow one to put an upper limit
  on the mass of the sterile neutrino. Unfortunately, such a computation is
  strongly model-dependent.  In~\cite{Dodelson:93}, \cite{Dolgov:00},
  \cite{Abazajian:01a}, \cite{Abazajian:05a}, the relic abundance of the
  sterile neutrinos was computed in a simple model with only one sterile
  neutrino, assuming the absence of the sterile neutrinos above the
  temperatures $\sim 1$ GeV. Yet, even the computation in this simplest model
  is subject to a number of
  uncertainties~\citep{Shi:98,Boyarsky:05,Boyarsky:06c,Asaka:06,Asaka:06b,
    Shaposhnikov:06}.  In particular, in~\cite{Dodelson:93}, \cite{Dolgov:00},
  \cite{Abazajian:01a}, \cite{Abazajian:05a}, two assumptions were made:
  \begin{inparaenum}[\em (i)]
  \item the absence of heavy particles, whose decay can dilute the relic
    abundance,
  \item the absence of lepton asymmetries.
    \end{inparaenum}
    In addition, simplifying assumptions about dynamics of hadrons at
    temperatures ${\cal O}(150)$ MeV were used.  Recently,~\cite{Asaka:06c}
    performed this computation from the first principles, showing that the
    uncertainty due to QCD effects (between minimal and maximal values of
    $\sin^2(2\theta)$ for given $M_s$) is about a factor of 8.
    
    Taking away the assumptions about the absence of the sterile neutrinos
    above the temperatures $\sim 1$ GeV makes any mixing angle possible.  For
    example, the DM neutrinos can be created due to the inflaton
    decay~\cite{Shaposhnikov:06}.  Therefore, in this work we chose not to
    derive an upper bound on the mass of the sterile neutrino.
\end{compactenum}

\begin{acknowledgements}
  
  We would like to thank G.~Gilmore, A.~Neronov, M.~Markevitch, I.~Tkachev,
  and M.~Shaposhnikov for help during the various stages of this project. J.N.
  acknowledges the support from the Academy of Finland. The work of A.B. was
  (partially) supported by the EU 6th Framework Marie Curie Research and
  Training network "UniverseNet" (MRTN-CT-2006-035863).  The work of O.R. was
  supported in part by European Research Training Network contract 005104
  ``ForcesUniverse'' and by a \emph{Marie Curie International Fellowship}
  within the $6^\mathrm{th}$ European Community Framework Programme.

\end{acknowledgements}

\appendix
\section{Determining parameters of NFW profile}
\label{sec:param-nfw}

Using the data on rotation curves, one usually obtains the following
parameters of DM distribution (see e.g. \cite{Klypin:02}): virial mass
$M_\vir$, virial radius $r_\vir$, and concentration parameter $C$. They have
the following relation with the parameters of NFW profile~(\ref{eq:6}) $r_s$
and $\rho_s$:
\begin{equation}
  \label{eq:7}
  r_s = \frac{r_\vir}C;\qquad \rho_s = \frac{M_\vir}{4\pi r_s^3 f(C)}\;,
\end{equation}
where in terms of function $f(x)$
\begin{equation}
  \label{eq:8}
  f(x) = \log(1+x)-\frac x{1+x}\;,
\end{equation}
one obtains the mass within the radius $r$:
\begin{equation}
  \label{eq:9}
  M(r) = M_\vir\frac{f(r/r_s)}{f(C)}\;.
\end{equation}
If DM distribution in the Milky Way is described by the NFW model (as in
\citealt{Battaglia:05,Klypin:02}), the flux from a direction $\phi$ is
given by
\begin{equation}
  \label{eq:12}
  F_\dm^\nfw(\phi) = \frac{\Gamma\Omega_\fov}{8\pi}\int\limits^\infty_0\hskip
  -.75ex dz\,
  \rho_\nfw(\sqrt{r_\odot^2 + z^2 + 2 r_\odot z \cos\phi})
\end{equation}
(notations are the same as in Eqs.(\ref{eq:19})--(\ref{eq:11})). Let us
consider two cases, when the integral in~(\ref{eq:12}) can be easily computed.
Namely,  we have for $\phi=180^\circ$
\begin{equation}
  \label{eq:13}
   F_\dm^\nfw(180^\circ)=\frac{\Gamma\Omega_\fov}{8\pi}\rho_s
   r_s\left[\log(1+\frac{r_s}{r_\odot}) -\frac{r_s}{r_s+r_\odot}\right]\;,
\end{equation}
and for $\phi=90^\circ$
\begin{equation}
  \label{eq:14}
  F_\dm^\nfw(90^\circ)  =\frac{\Gamma\Omega_\fov}{8\pi}\rho_s r_s\left[-1-\log\frac{2r_s}{r_\odot}\right.
     \quad +\left.\frac{r_\odot^2}{r_s^2}\left(\frac32\log\frac{2r_s}{r_\odot} -\frac54\right)\right] + \mathcal{O}\parfrac{r_\odot^4}{r_s^4}
\end{equation}
(in the latter case the analytic expression is too complicated, so we present
Taylor expansion for the case $r_\odot\ll r_s$).


\end{document}